\documentstyle[preprint,eqsecnum,aps]{revtex}
\begin{document}
\draft
\title{Canonical Quantization of the Maxwell-Chern-Simons Theory in the Coulomb
Gauge\cite{byline}}
\author{F. P. Devecchi, M. Fleck, H. O. Girotti}
\address{Instituto de F\'{\i}sica,
Universidade Federal do Rio Grande do Sul \\ Caixa Postal 15051, 91501-970  -
Porto Alegre, RS, Brazil.}
\author{M. Gomes and A. J. da Silva}
\address{Instituto de F\'{\i}sica, Universidade de S\~ao Paulo, \\ Caixa
Postal 20516, 01452--990, S\~ao Paulo, SP, Brazil.}
\date{August 1994}
\maketitle

\begin{abstract}
The Maxwell-Chern-Simons theory is canonically quantized in the Coulomb gauge
by using the Dirac bracket quantization procedure. The determination of the
Coulomb gauge polarization vector turns out to be intrincate. A set of quantum
Poincar\'e densities obeying the Dirac-Schwinger algebra, and, therefore, free
of anomalies, is constructed. The peculiar analytical structure of the
polarization vector is shown to be at the root for the existence of spin of
the massive gauge quanta.The Coulomb gauge Feynman rules are used
to compute the M\"oller scattering amplitude in the lowest order of
perturbation theory. The result coincides with that obtained by using covariant
Feynman rules. This proof of equivalence is, afterwards, extended to all
orders of perturbation theory. The so called infrared safe photon propagator
emerges as an effective propagator which allows for replacing all the terms in
the interaction Hamiltonian of the Coulomb gauge by the standard field-current
minimal interaction Hamiltonian.
\end{abstract}

\pacs{PACS: 11.10.Kk, 11.10.Ef}
\narrowtext

\section{Introduction}
\label{sec:level1}
As is known\cite{Ja,Sch}, the Maxwell-Chern-Simons (MCS) theory is a
$(2+1)$-dimensional field model describing the coupling of charged fermions
$(\bar{\psi},\psi)$ of mass $m$ and electric charge $e$ to the electromagnetic
potential $A_{\mu}$ via the Lagrangian density

\begin{equation}
\label{11}
{\cal L} = -\frac{1}{4}F_{\mu\nu}F^{\mu\nu}+\frac{\theta}{4}
\epsilon^{\mu\nu\alpha}F_{\mu\nu}A_{\alpha}
+\frac{i}{2}\bar{\psi}\gamma^{\mu}\partial_{\mu}\psi-
\frac{i}{2}(\partial_{\mu}\bar{\psi})\gamma^{\mu}\psi-m\bar{\psi}\psi
+e\bar{\psi}\gamma^{\mu}A_{\mu}\psi ,
\end{equation}

\noindent
where $F_{\mu\nu}\equiv\partial_{\mu}A_{\nu}-\partial_{\nu}A_{\mu}$ and
$\theta$ is a parameter with dimension of mass.
Neither parity nor time reversal are, separately,
symmetries of the model \footnote{Throughout this paper we use natural units
$(c=\hbar=1)$. Our metric is $g_{00}=-g_{11}=-g_{22}=1$. For the
$\gamma$-matrices we adopt the representation
$\gamma^{0}=\sigma^{3},\gamma^{1}=i\sigma^{1},\gamma^{2}=i\sigma^{2}$, where
$\sigma^{i},i=1,2,3$ are the Pauli spin matrices. The fully antisymmetric
tensor $\epsilon^{\mu \nu \lambda}$ is normalized such that
$\epsilon^{012}=1$ and we define $\epsilon^{ij} \equiv \epsilon^{0ij}$.
Repeated greek indices sum from 0 to 2, while repeated latin indices from the
middle of the alphabet sum from 1
to 2.}

The quantization of the MCS model in covariant gauges is free of
inconsistencies\cite{Ja}. On the other hand, the quantization of the free MCS
theory in the Coulomb gauge already demands special care, due to the
appearance of
infrared divergences\cite{Ja}. Although it has been argued\cite{Ha} that these
singularities can be avoided and some results of the Coulomb gauge version
of the theory have been used from time to time for different purposes\cite{Fo},
the consistency of this formulation has not yet been fully
established. This paper is dedicated to study the Coulomb gauge quantization
of the MCS model.

In Section II we determine the polarization vector for the free MCS theory in
the Coulomb gauge. This is an essential piece of information in what follows
and, as will be seen, it will be done by gauge transforming the polarization
vector of the Landau gauge.

In Section III the free MCS theory is canonically quantized, in the Coulomb
gauge, by using the Dirac bracket quantization procedure\cite{Di,Fr,Su,Gi}.
After finding the equal-time commutation relations
and the canonical Hamiltonian, we build the reduced phase space\cite{Fr,Gi}
and, then, solve the Heisenberg equations of motion for the independent
variables. Afterwards, the Hilbert space of physical states is constructed.
As we shall see, all excitations turn out to be massive.

Section IV is dedicated to study the Poincar\'e symmetry for the free MCS
model. We first show that there exist a set of densities with vanishing vacuum
expectation values and obeying the Dirac-Schwinger equal-time commutator
algebra\cite{Sch1}. These densities are, therefore, free of anomalies and
their space integrals yield the corresponding Poincar\'e generators. Of
particular interest is the generator of rotations because of the delicate
mechanism giving rise to the spin of the gauge particle.

The fermions are brought into the game in Section V. The Dirac bracket
quantization procedure is used again
to quantize the full theory in the Coulomb gauge. Then, the free Hamiltonian,
the interaction Hamiltonian and the Feynman rules are found. The core of
the Section is concerned with the computation of the lowest order contribution
to the M\"oller scattering amplitude by using the Coulomb gauge Feynman rules.
We demonstrate that the result agrees with that found by
using covariant gauge Feynman rules\cite{Gi1,Gi2} and, moreover, that the
infrared safe propagator\cite{Ja,Lo} arises as an effective photon propagator
which allows for replacing all non-covariant terms in the interaction
Hamiltonian by the field-current minimal interaction

In Section VI we analyze the M\"oller scattering amplitude to all orders in
perturbation theory. We prove, order by order,
that the Coulomb and the covariant gauge Feynman rules yield the same result.

The conclusions are contained in Section VII.

\section{Determination of the Polarization Vector in the Coulomb Gauge}
\label{sec:level2}

The dynamics of the free MCS theory is described by the Lagrangian density

\begin{equation}
\label{21}
{\cal L} = -\frac{1}{4}F_{\mu\nu}F^{\mu\nu}+\frac{\theta}{4}
\epsilon^{\mu\nu\alpha}F_{\mu\nu}A_{\alpha},
\end{equation}

\noindent
from which one derives the following equation of motion for the field
$A^{\beta}$,

\begin{equation}
\label{22}
\Box  A^{\beta }-\partial ^{\beta }( \partial _{\alpha } A^{\alpha })
+\theta \epsilon ^{\beta \mu \alpha }\partial _{\mu } A_{\alpha} = 0.
\end{equation}

\noindent
After taking into account the Coulomb gauge condition,

\begin{equation}
\label{23}
\chi\, \equiv \,\partial^{j}A^{j} = 0,
\end{equation}

\noindent
the equation of motion (\ref{22}) reduces to

\begin{equation}
\label{24}
\Box  A^{\beta }-\partial ^{\beta }( \partial _{0} A^{0})
+\theta \epsilon ^{\beta \mu \alpha }\partial _{\mu } A_{\alpha} = 0.
\end{equation}

\noindent
In Hamiltonian language, the MCS theory possesses two first-class constraints
and, hence, two subsidiary conditions are needed to fix the gauge
completely\cite{Fo}. Thus, one is left with only two independent variables in
phase space, one coordinate and one momentum. Correspondingly, this theory only
exhibits one degree of freedom in configuration space\cite{Mo}. Therefore,
it should be possible to write a plane wave solution of (\ref{24}) in terms
of a single polarization vector ($\epsilon^{\beta}(\vec{k})$), namely,

\begin{equation}
\label{25}
A^{\beta}(x)\,=\,\epsilon^{\beta}(k) e^{ik\cdot x}.
\end{equation}

\noindent
When (\ref{25}) is inserted back into (\ref{23}) and (\ref{24}) one obtains,
respectively,
\begin{equation}
\label{26}
k^{i}\epsilon^{i}(k)\,=\,0
\end{equation}

\noindent and

\begin{equation}
\label{27}
\Sigma^{\beta \alpha} \epsilon_{\alpha}(k)\,=\,0\,,
\end{equation}

\noindent
where

\begin{equation}
\label{271}
\Sigma^{\beta \alpha} \equiv - k^{2} g^{\beta \alpha} \,
+\,k^{\beta} k^{0} g^{0 \alpha}\,
+\,i \theta \epsilon^{\beta \rho \alpha} k_{\rho}.
\end{equation}

\noindent
The vanishing of the determinant of the matrix $\Sigma^{\beta \alpha}$ is a
necessary and sufficient condition for the homogeneous system of equations in
(\ref{27}) to have solutions different from the trivial one
$\epsilon^{\beta}(k)=0$. One easily finds that

\begin{equation}
\label{272}
\det \Sigma\,=\,|{\vec{k}}^2|\, k^2 \,(k^2 - \theta^2).
\end{equation}

\noindent
Hence, there are, in principle, three independent solutions for the system
(\ref{27}). This
seems to contradict the above conclusion based on the counting of the number of
independent degrees of freedom. However, one is to notice that the polarization
vector associated with the massless mode $k^2 = 0$ can only occur if
$k^{\beta}=0$ which in turn implies that $A^{\beta}$ is just a constant. To the
same conclusion one arrives by specializing (\ref{27}) and (\ref{271}) to the
case ${\vec{k}}^2 = 0$. Thus, as previously asserted, only one of the
excitations is dynamical. This is the one associated with the massive mode
$k^2 = \theta^2$.

We next focus on finding the polarization vector for the massive excitation
$k^2 = \theta^2$. The transversality condition (\ref{26}) is readily satisfied
by choosing

\begin{equation}
\label{28}
\epsilon^{i}(\vec{k})\,=\,\epsilon^{ij}\,k^{j}\,b({\vec{k}}^{2}).
\end{equation}

\noindent
With the aim of determining the unknown function $b({\vec{k}}^{2})$, we replace
(\ref{28}) into (\ref{27}). For $\beta=0$ we obtain

\begin{equation}
\label{29}
\epsilon^{0}(\vec{k})\,=\,i\, \theta \,b({\vec{k}}^{2}),
\end{equation}

\noindent
while for $\beta=i$ one arrives at

\begin{equation}
\label{201}
(-k^{2}\,+\,\theta^{2})\epsilon^{ij}\,k^{j}\,b({\vec{k}}^{2})\,=\,0,
\end{equation}

\noindent
which sais that the function $b({\vec{k}}^{2})$ remains unknown. One may argue
that a normalization condition for the space-like vector $\epsilon^{\mu}(k)$,
that is still lacking, is all what is needed to solve for $b(\vec{k})$.
However, through this kind of device one only determines the modulus of
$b(\vec{k})$, while a possible phase factor would
be missed. As will be seen, it is precisely a $k$-dependent phase factor which
accounts for the existence of spin in the present case. Furthermore, a simple
calculation shows that, for example, $\epsilon^{\mu}(k)
\epsilon_{\mu}^{\ast}(k)=-1$ leads to a function $|b(\vec{k})|$ which does not
fulfills the regularity requirements assumed for a polarization vector.
This difficulties recognize as common origin the fact that one is dealing with
massive gauge quanta. Indeed, $\theta \neq 0$ implies in the existence of a
rest frame of reference ($\vec{k}=0$) for these particles and, in such frame,
the Coulomb condition (\ref{26}) becomes ambiguous. This is an entirely new
situation as compared with that encountered, for instance, in the Coulomb
gauge formulation of QED$_4$, where the gauge particle is massless.

We are, then, forced to adopt a new strategy for finding for $b({\vec{k}}^2)$.
It consists in reaching the Coulomb
gauge from the Landau gauge ($\partial_{\mu}A^{\mu}_L=0$) through the gauge
transformation linking these two gauges\footnote{The gauge
subscript $L$ identify quantities belonging to the Landau gauge.
Coulomb gauge quantities remain without gauge identification.}. One is to
observe that, unlike the case of the Coulomb gauge, the Landau gauge condition

\begin{equation}
\label{202}
k_{\beta}\epsilon^{\beta}_L(k)\,=\,0
\end{equation}

\noindent
remains operative even in the rest frame of reference ($\vec{k}=0$). In Landau
gauge, (\ref{271}) is replaced by

\begin{equation}
\label{2021}
\Sigma^{\beta \alpha}_L \equiv - k^{2} g^{\beta \alpha}\,
+\,i \theta \epsilon^{\beta \rho \alpha} k_{\rho}\,,
\end{equation}

\noindent
whose determinant is

\begin{equation}
\label{2022}
\det \Sigma _{L}\,=\, k^2 (k^2 - \theta^2)\,.
\end{equation}

\noindent
As seen, there are massless and massive excitations in the Landau
gauge\cite{Jap1}.

For $k^2 = 0$, the simultaneous solving of (\ref{202}) and

\begin{equation}
\label{203}
\Sigma^{\beta \alpha}_L \epsilon_{L,\alpha}(k)\,=\,0\,
\end{equation}

\noindent
is straightforward and yields

\begin{equation}
\label{2031}
\epsilon^{\beta}_L(k) = k^{\beta} \eta (k) \,,
\end{equation}

\noindent
where $\eta (k)$ is an arbitrary function of $k$. This confirms that the
massless excitations are pure gauge artefacts\cite{Jap2}.

On the other hand, for $k^2 = \theta^2$ one finds that

\begin{mathletters}
\label{204}
\begin{eqnarray}
\epsilon^{0}_L (\vec{k})\,& = &\, \frac{1}{|\theta|}\,
\vec{k} \cdot {\vec{\epsilon}}_{L}(0), \label{mlett:a204} \\
\epsilon^{i}_L (\vec{k})\,& = &\,\epsilon^{i}_L(0)\,+\,
\frac{{\vec{\epsilon}}_{L}(0)
\cdot \vec{k}}{(\omega_{\vec{k}}\,+\,|\theta|)\,|\theta| }
\,k^{i}, \label{mlett:b204}
\end{eqnarray}
\end{mathletters}

\noindent
where $\epsilon^{\beta}_L(0)$ is the polarization in the rest frame of
reference and $\omega_{\vec{k}}\equiv+(\vec{k}^{2}+\theta^{2})^{1/2}$.
{}From Eq.(\ref{204}) follows that\cite{Bo}

\begin{mathletters}
\label{205}
\begin{eqnarray}
\epsilon^{0}_L(0)\,& = &\,0, \label{mlett:a205} \\
\epsilon^{2}_L(0)\,& = &\,-i\, \frac{\theta}{|\theta|}\,\epsilon^{1}_L(0).
\label{mlett:b205}
\end{eqnarray}
\end{mathletters}

\noindent
To summarize, the physically meaningful Landau gauge polarization vector
$\epsilon^{\beta}_L(\vec{k})$ turns out to be a complex space-like vector
obeying

\begin{mathletters}
\label{206}
\begin{eqnarray}
\epsilon^{\beta}_L(\vec{k})\epsilon_{L,\beta}(\vec{k})
\,& = &\,-{\vec{\epsilon}}_{L}(0)\cdot{\vec{\epsilon}}_{L}(0)\,=\,0,
\label{mlett:a206} \\
\epsilon^{\beta}_L(\vec{k})\epsilon^{\ast}_{L,\beta}(\vec{k})
\,& = &\,-{\vec{\epsilon}}_{L}(0)\cdot{\vec{\epsilon}}^{\ast}_L(0)\,=\,
-2\,|\epsilon^{1}_L(0)|^{2}\label{mlett:b206},
\end{eqnarray}
\end{mathletters}

\noindent
where $|\epsilon^{1}_L(0)|$ is to be fixed by normalization.

We look next for the gauge transformation $\Lambda(x)$ linking the Landau and
Coulomb gauges. It must be such that

\begin{equation}
\label{207}
\epsilon^{\beta}(k)\,=\,\epsilon^{\beta}_L(k)\,
+\,i\,k^{\beta}\,\lambda(k),
\end{equation}

\noindent
where $\Lambda(x)=\lambda(k)\exp{ik\cdot x}$. Irrespective of whether one is
dealing with the massless or with the massive mode, one finds from (\ref{207})

\begin{equation}
\label{2071}
\lambda(k)\,=\,i\,\frac{\vec{k} \cdot {\vec{\epsilon}}_{L}(\vec{k})}
{{\vec{k}}^2}.
\end{equation}

According to (\ref{2031}), $\lambda(k)$ reduces, in the case of the massless
mode, to $\lambda(k) = i\eta(k)$. When this result is replaced back into
(\ref{207}) one gets $\epsilon^{\beta}(k) = 0$ indicating that the massless
(pure gauge) excitations are not present in the
Coulomb gauge. As in 3+1-dimensions, the Coulomb gauge remains a faithful
gauge in the sense of containing only physical excitations.

In the case of massive excitations one is to replace (\ref{204}) into
(\ref{2071}), thus obtaining

\begin{equation}
\label{208}
\lambda(\vec{k})\,=\,i \frac{{\vec{\epsilon}}_{L}(0)
\cdot \vec{k}} {{\vec{k}}^{2}}\,+i\,\frac{{\vec{\epsilon}}_{L}(0)
\cdot \vec{k} }{(\omega_{\vec{k}}\,+\,|\theta|)\,|\theta| }.
\end{equation}

\noindent
{}From Eqs.(\ref{204}), (\ref{207}) and (\ref{208}) one now finds that

\begin{mathletters}
\label{209}
\begin{eqnarray}
\epsilon^{0}(\vec{k})\,& = & \, i\,\theta\,\frac{\epsilon^{i}_L(0)\,
\epsilon^{ij} k^{j}} {{\vec{k}}^{2}}, \label{mlett:a209} \\
\epsilon^{i}(\vec{k})\,& = & \, \left(\delta^{ij} - \frac{k^i k^j}{|\vec{k}|^2}
\right) \epsilon^{j}_L(0)\,=\,
\left[ i\,\frac{\theta}{|\theta|}\,\epsilon^{1}_L(0)\right]\,
\frac{\epsilon^{ij} k^{j}}{|\vec{k}|} e^{-i\frac{\theta}{|\theta|} \phi},
\label{mlett:b209}
\end{eqnarray}
\end{mathletters}

\noindent
where

\begin{equation}
\label{210}
\phi\, \equiv \,\arctan{\frac{k^{2}}{k^{1}}}
\end{equation}

\noindent
is the angle between the spatial vectors $\vec{k}$ and
${\vec{\epsilon}}_{L}(0)$. In Eqs.(\ref{209}) we quoted the final form of the
components of the Coulomb gauge polarization vector. We succeded in expressing
$\epsilon^{\beta}(\vec{k})$ in a form that makes clear that it goes
continuously to the corresponding value in the rest frame of reference. The
peculiar structure of $ \epsilon^{i}(\vec{k}) $ (see Eq.(\ref{mlett:b209}))
should be observed. It is at the origin of the spin of the massive gauge
quanta.

\section{Canonical Quantization of the free MCS theory in the Coulomb Gauge}
\label{sec:level3}

Within the Hamiltonian framework the free MCS theory is, as already pointed
out\cite{Fo}, fully characterized by the canonical Hamiltonian
\begin{equation}
\label{32}
H = \int d^2z \left[ \frac{1}{2}\pi_j\pi_j-\frac{\theta}{2}\pi_i\epsilon ^{ik}
A^k+\frac{1}{4}F^{ij}F^{ij}+\frac{\theta ^2}{8}A^jA^j \right]
\end{equation}

\noindent
the primary first-class constraint

\begin{equation}
\label{33}
\Omega_{0}\, \equiv \,\pi_{0}\, \approx \,0
\end{equation}

\noindent
and the secondary firs-class constraint

\begin{equation}
\label{34}
\Omega_{1}\, \equiv \partial^{i}\pi_{i}\,+\,\frac{\theta}{2}
\epsilon^{ij} \partial^{i} A^{j} \, \approx \,0.
\end{equation}

\noindent
Here, we have designated by $\pi_{\mu}$ the momenta canonically conjugate to
$A^{\mu}$. As usual, the sector $A^{0}, \pi_{0}$ can be eliminated from the
phase space; $\pi_{0}$ is fixed by the constraint condition (\ref{33}) while
$A^{0}$ acts as the Lagrange multiplier of $\Omega_{1}$ and will be
determined, after gauge fixing, as a function of the remaining canonical
variables.

The quantization of the system in the Coulomb gauge,
$\chi \equiv \partial^{j}A^{j}=0$, is straightforward. The set of constraints
$\Omega_{1}\approx 0$ and $\Omega_{2} \equiv \chi \approx 0$ is, by
construction, second class and Dirac brackets\cite{Di} can be introduced in
the usual manner. One then promotes the phase space variables $A^{i},\pi_{i}$
to selfadjoint operators\footnote{To simplify the notation, we shall not
distinguish between a quantum field operator and its classical counterpart},
and establishes that these operators are to obey a set of equal-time
commutation rules which are abstracted from the corresponding Dirac brackets,
the constraints and gauge conditions thereby translating into strong operator
relations. This is the Dirac bracket quantization procedure\cite{Di,Fr,Su,Gi},
which presently yields

\begin{mathletters}
\label{35}
\begin{eqnarray}
& &[A^{i}(x^{0},\vec{x})\,,\,A^{j}(x^{0},\vec{y})]\,=\,0, \label{mlett:a35} \\
& &[A^{i}(x^{0},\vec{x})\,,\,\pi_{j}(x^{0},\vec y)]\,
=\,i\,P^{ij}_{T}(\vec{x})\,\delta(\vec{x}-\vec{y}), \label{mlett:b35} \\
& &[\pi_{i}(x^{0},\vec{x})\,,\,\pi_{j}(x^{0},\vec{y})]\,
=\,-i\,\frac{\theta}{2}\,\epsilon^{ij}\,\delta(\vec{x}-\vec{y}),
\label{mlett:c35}
\end{eqnarray}
\end{mathletters}

\noindent
where $P^{ij}_{T}(\vec{x})\equiv
\delta^{ij}-\partial^{i}_{x}\partial^{j}_{x}/\nabla^{2}_{x}$ and
$\nabla^{2}_{x}\equiv \partial^{i}_{x}\partial^{i}_{x}$.

We look next for the reduced phase-space, i.e., the phase-space spanned by the
independent variables. It will prove convenient, for this purpose, to split
$A^{i}(x)$ and $\pi_{i}(x)$ into longitudinal and transversal ($T$)
components. From $\Omega_{2}\approx 0$ follows that the longitudinal component
of $A^{i}(x)$ vanishes, whereas $\Omega_{1}\approx 0$ can be used to eliminate
the longitudinal component of $\pi_{i}(x)$ in terms of $A^{i}_{T}(x)$. As
consequence, the theory can be fully phrased in terms of $A^{i}_{T}(x)$ and
$\pi_{i}^{T}(x)$. It is not difficult to check that the canonical Hamiltonian
(\ref{32}) and the equal-time commutation rules (\ref{35}), when casted in
terms of the independent variables, read, respectively, as follows

\begin{equation}
\label{38}
H\,=\,\int d^{2} z [\frac{1}{2}\pi ^T_j\pi^T_j
- \frac {\theta}{2}\pi^T_j\epsilon^{jk}
A^k_T+\frac{1}{4}F^{ij}F^{ij}+\frac{\theta^2}{2}A^j_TA^j_T],
\end{equation}

\begin{mathletters}
\label{39}
\begin{eqnarray}
& &[A^i_T(x^0,\vec x)\,,\,A^j_T(x^0,\vec y)]\,= \,0 , \label{mlett:a39} \\
& &[A^i_T(x^0,\vec x)\,,\,\pi^T_j(x^0,\vec y)]\,
= \,i\,\, P^{ij}_T(\vec x)\,\delta(\vec x-\vec y), \label{mlett:b39} \\
& &[\pi ^T_i(x^0,\vec x)\,,\,\pi ^T_j(x^0,\vec y)]\,=\,0. \label{mlett:c39}
\end{eqnarray}
\end{mathletters}

The Heisenberg equations of motion deriving from (\ref{38}) and (\ref{39})
are

\begin{mathletters}
\label{301}
\begin{eqnarray}
& &\partial^0 A^i_T(x^0,\vec x)\,=\,\pi^T_i(x^0,\vec x), \label{mlett:a301} \\
& &\partial_0\pi^T_i(x^0,\vec x)\,=\,\partial^j F^{ji}(x^0,\vec x)
-\theta^2 A^i_T(x^0,\vec x), \label{mlett:b301}
\end{eqnarray}
\end{mathletters}

\noindent
which, after decoupling, yield

\begin{equation}
\label{302}
(\Box + \theta^2)A^i_T(x^0,\vec{x})\,=\,0,
\end{equation}

\noindent
showing that the excitations are all massive. As one can easily verify, the
field configuration

\begin{equation}
\label{303}
A^{i(\pm)}_{T}(x^{0},\vec{x})\,=\,\frac{1}{2\pi} \int
\frac{d^{2}k}{\sqrt{2\omega_{\vec{k}}}} e^{\pm i(\omega_{\vec{k}} x^{0}-\vec{k}
\cdot \vec{x})}\,A^{i(\pm)}_{T}(\vec{k}),
\end{equation}
solves the equation of motion (\ref{302}) and the equal-time commutation
relations (\ref{39}). Here, we have introduced the definitions

\begin{equation}
\label{304}
A^{i(\pm)}_{T}(\vec{k})\,\equiv\, \epsilon^{i}(\vec{k})\,a^{\pm}(\vec{k}),
\end{equation}

\noindent
where $\epsilon^{i}(\vec{k})$ is the Coulomb gauge polarization vector, found
in the previous Section, and

\begin{mathletters}
\label{305}
\begin{eqnarray}
& &[a^{-}(\vec{k})\,,\,a^{-}(\vec{k\prime})]\,
=\,[a^{+}(\vec{k})\,,\,a^{+}(\vec{k\prime})]\,=\,0, \label{mlett:a305} \\
& &[a^{-}(\vec{k})\,,\,a^{+}(\vec{k\prime})]\,=\,
\frac{1}{{|\epsilon^{1}(0)|}^{2}}\,\delta(\vec{k}-\vec{k\prime}).
\label{mlett:b305}
\end{eqnarray}
\end{mathletters}

\noindent
As known, the creation-destruction algebra (\ref{305}) can be implemented in a
Hilbert space with positive definite metric. So far, the canonical quantization
of the MCS theory in the Coulomb gauge does not appear to be afflicted by
ambiguities and can be sistematically implemented. In the next section we shall
investigate the Poincar\'e symmetry of the model and the origin of the spin of
the gauge particles.

\section{Poincar\'e Invariance of the free MCS Theory. The Spin of the MCS
quanta}
\label{sec:level4}

We start by considering the normally ordered composite operators

\begin{mathletters}
\label{41}
\begin{eqnarray}
& &\Theta^{00}(x)\,\equiv\, :F^{0k}(x)F^{0k}(x):\,
+\,\frac {1}{4} : F^{\mu \nu}(x) F_{\mu \nu }(x):\,, \label{mlett:a41} \\
& &\Theta^{0k}(x)\,\equiv\, :F^{0j}(x)F^{kj}(x):\,, \label{mlett:b41} \\
& &\Theta^{kj}(x)\,\equiv\,-:F^k_{\lambda}(x)F^{j\lambda}(x):\,
- \,\frac{1}{4}\delta^{kj}
:F^{\mu \nu}(x)F_{\mu \nu}(x):\,. \label{mlett:c41}
\end{eqnarray}
\end{mathletters}

\noindent
Since we are working in the Coulomb gauge, the space components of the vector
$A^{\mu}(x)$ are purely transversal while the time component $A^{0}(x)$ is, as
we already said, a Lagrange multiplier given in terms of the remaining
variables by the expression

\[
A^{0}(x)\,=\,\frac{\theta}{\nabla^{2}_{x}}\,\epsilon^{ij} \partial^{i}
A^{j}_{T}(x) .
\]

\noindent
Furthermore, all the velocities in (\ref{41}) can be eliminated in favor of
the momenta by using the Heisenberg equation of motion (\ref{mlett:a301}).
Thus, all the composite operators defined in (\ref{41}) can be entirely written
in terms of the independent phase-space variables. By using (\ref{39}) one can
check, afterwards, that these operators indeed verify the Dirac-Schwinger
algebra\cite{Sch1},

\begin{mathletters}
\label{42}
\begin{eqnarray}
& &[\Theta^{00}(x^{0}, \vec{x})\,,\,\Theta^{00}(x^{0}, \vec{y})]\,
=\,-i\,\left(\Theta^{0k}(x^{0}, \vec{x})\,
+\,\Theta^{0k}(x^{0},\vec{y})\right) \partial^{x}_{k}
\delta(\vec{x}-\vec{y}), \label{mlett:a42} \\
& &[\Theta^{00}(x^{0}, \vec{x})\,,\,\Theta^{0k}(x^{0},\vec{y})]\,=\,
-i\,\left(\Theta^{kj}(x^{0},\vec{x})\,
-\,g^{kj}\,\Theta^{00}(x^{0},\vec{y})\right)
\partial^{x}_{j}\delta(\vec{x}-\vec{y}), \label{mlett:b42} \\
& &[\Theta^{0k}(x^{0},\vec{x})\,,\,\Theta^{0j}(x^{0},\vec{y})]\,=\,-i\,
\left(\Theta^{0k}(x^{0},\vec{x})\partial_{j}^{x}\,+\,
\Theta^{0j}(x^{0},\vec{y})\partial^{x}_{k} \right) \delta(\vec{x}-\vec{y}),
\label{mlett:c42}
\end{eqnarray}
\end{mathletters}

\noindent
and can, therefore, be taken as the Poincar\'e densities of the free MCS
theory. The generators of space-time translations ($P^{\mu}$), Lorentz boosts
($J^{0i}$) and spatial rotations ($J$) are defined in the standard manner,
\begin{mathletters}
\label{43}
\begin{eqnarray}
& &P^{0}\,\equiv\,\int d^2x\,\Theta^{00}(x^{0},\vec{x}) \,=\,H,
\label{mlett:a43} \\
& &P^{i}\,\equiv\,\int d^2x\,\Theta^{0i}(x^{0},\vec{x}),
\label{mlett:b43} \\
& &J^{0i}\,\equiv\,- x^{0}P^{i}\,+\,\int d^2x\,
[x^{j}\Theta^{00}(x^{0},\vec{x})], \label{mlett:c43} \\
& &J\,\equiv\,\epsilon^{ij} \int d^{2}x \,x^{i} \Theta^{0j}(x^{0},\vec{x}),
\label{mlett:d43}
\end{eqnarray}
\end{mathletters}

\noindent
and can be seen to fulfill the Poincar\'e algebra. We stress that, within the
present formulation, the commutator $[J^{0i},J^{0k}]$,

\begin{equation}
\label{44}
[J^{0i}\,,\,J^{0k}]\,=\,-i\,\epsilon^{ik} J,
\end{equation}

\noindent
is free of anomalies\cite{Ja}.

We next analyze the spin content of the MCS quanta. By going with
(\ref{303}) into Eq.(\ref{mlett:b41}) and with the result thus obtained into
(\ref{mlett:d43}) one finds that

\begin{equation}
\label{45}
J\,=\,i\epsilon^{jl}\int
d^{2}k\,A^{m(+)}_{T}(\vec{k})\,k^{l} \, \frac{\partial}{\partial
k^{j}}\,A^{m(-)}_{T}(\vec{k})\,,
\end{equation}

\noindent
which after taking into account (\ref{304}) and the explicit form of the
Coulomb gauge polarization vector derived in Section II (see
Eq.(\ref{mlett:b209})) can be casted as

\begin{equation}
\label{46}
J\,=\,{|\epsilon^{1}(0)|}^{2}\,\frac{\theta}{|\theta|} \int d^{2}k\,
a^{+}(\vec{k})
a^{-}(\vec{k})\,+\,i \epsilon^{jl} {|\epsilon^{1}(0)|}^{2} \int d^{2}k\,
a^{+}(\vec{k})\,
k^{l} \frac{\partial}{\partial k^{j}}\, a^{-}(\vec{k}).
\end{equation}

\noindent
The first term in the right hand side of (\ref{46}) is the spin part of the
total angular momentum. It originates from the exponential in
$(\ref{mlett:b209})$. The action of the operator $J$ on a single particle
state ($a^{+}(\vec{k})|0>$) can be readily derived. In particular, for the
rest frame of reference one obtains

\begin{equation}
\label{47}
J\,\{ a^{+}(\vec{k} = 0)|0> \} \,
=\,\frac{\theta}{|\theta|} \{ a^{+}(\vec{k} = 0)|0> \},
\end{equation}

\noindent
which tell us that the spin of the MCS quanta is $\pm 1$ depending upon the
sign of the topological mass factor\cite{Ja,Bi}.

\section{Lowest Order M\"oller Scattering Amplitude in the MCS theory}
\label{sec:level5}

In this Section we bring the fermions into the game. Within the Hamiltonian
approach the dynamics of the MCS model in the presence of fermions is described
by the canonical Hamiltonian

\begin{eqnarray}
\label{51}
&& H^{F}\, =  \int d^2z \,\left[ \frac {1}{2} \pi_j\pi_j\,
-\,\frac{\theta }{2}\pi_i \epsilon^{ik} A^k\,+\, \frac {1}{4}F^{ij} F^{ij}
+\frac {\theta^2}{8} A^j A^j \right. \nonumber \\
&& \left. + \frac{1}{2} \pi_{\psi} \cdot \gamma^{0} \gamma^{j} \partial^{j}
\psi\,+\,\frac{1}{2} (\partial^{j} \pi_{\psi}) \cdot \gamma^{0} \gamma^{j} \psi
\,-\,i e \,\pi_{\psi} \gamma^0 \gamma^j \psi A^j\,
-\,i m \pi_{\psi} \gamma^{0} \psi \right],
\end{eqnarray}

\noindent
the primary first-class constraint (\ref{33}) and the secondary first
constraint

\begin{equation}
\label{52}
\Omega^{F}_{2}(x)=\partial^j\pi_j(x)\,+\,\frac{\theta}{2}\epsilon^{ij}
\partial^{i} A^{j}(x)\,-\,ie\pi_{\psi}(x) \cdot \psi(x)\approx 0.
\end{equation}

\noindent
Here, $\pi_{\psi}(x) = i \bar{\psi}(x)\gamma^{0} $ is the momentum canonically
conjugate to the field variable $\psi(x)$ and the dot symbolizes the
antisymmetrization  prescription ($2 A\cdot B\equiv  AB - BA$). The
quantization in the Coulomb gauge is performed, as in the case of the free MCS
theory, by means of the Dirac bracket quantization procedure. For the bosonic
sector one gets again the commutation relations in Eq.(\ref{35}), while the
equal-time commutation relations involving the fermionic field variables are

\begin{mathletters}
\label{53}
\begin{eqnarray}
& &[\pi_j (x^{0}, \vec{x})\,,\,\psi(x^{0}, \vec{y})]\,=\,
e \left[ \frac{\partial_j^x}{\nabla^{2}_{x}} \delta (\vec{x}-\vec{y}) \right]
\psi(x^{0}, \vec{y})\,, \label{mlett:a53} \\
& &[\pi_j (x^{0}, \vec{x})\,,\,\pi_{\psi}(x^{0}, \vec{y})]\,=\,
- e \left[ \frac{\partial_j^x}{\nabla^{2}_{x}} \delta (\vec{x}-\vec{y}) \right]
\pi_{\psi}(x^{0}, \vec{y})\,, \label{mlett:b53} \\
& &\{\psi(x^{0}, \vec{x})\,,\,\pi_{\psi}(x^{0}, \vec{y})\}\,=\,i\,\delta
(\vec{x}-\vec{y}). \label{mlett:c53}
\end{eqnarray}
\end{mathletters}

The reduced phase-space is now spanned by the independent variables
$A_{T}^{i}$,  $\pi^{T}_{i}$, $\psi$ and $\pi_{\psi}$. In term of these
variables the Hamiltonian $H^{F}$ splits into the sum of a free
($H_{0}$) plus an interacting part ($H_{I}$), i.e.,

\begin{equation}
\label{54}
H^{F}\,=\,H_{0}\,+\,H_{I}\,,
\end{equation}

\noindent
where

\begin{eqnarray}
\label{55}
H_0\,& = & \,\int d^2z\left[\frac {1}{2}\pi_j^T \pi^T_j\,
-\,\frac {\theta}{2} \pi_i^T
\epsilon^{ij} A^{j}_T\,+\,\frac {1}{4}F^{ij}_{T}F^{ij}_T\,
+\,\frac {\theta^2}{2} A^j_T A^j_T\ \right.  \nonumber \\
&& \left. + \frac{1}{2} \pi_{\psi} \cdot \gamma^0
\gamma^j \partial^j \psi\,+\,\frac{1}{2} (\partial^{j} \pi_{\psi}) \cdot
\gamma^{0} \gamma^{j} \psi\,-\, i m \pi_{\psi} \cdot \gamma^0 \psi \right]\,
\end{eqnarray}

\noindent
and

\begin{equation}
\label{56}
H_I=\int d^2z\left[ -ie \pi_{\psi} \cdot \gamma^0 \gamma^j A^j_T \psi\,
+\,i e \theta \pi_{\psi}\cdot \psi \epsilon^{ij} \frac{\partial^i}{\nabla^2}
A^{j}\,+\,\frac {e^2}{2}
(\pi_\psi \cdot \psi )\frac {1}{\nabla^2}(\pi_{\psi} \cdot \psi) \right]\,.
\end{equation}

\noindent
On the other hand, the only nonvanishing (anti)commutators turn out to be

\begin{mathletters}
\label{57}
\begin{eqnarray}
& &[A^i_T(x^0,\vec x)\,,\,\pi^T_j(x^0,\vec y)]\,
= \,i\,\, P^{ij}_T(\vec x)\,\delta(\vec x-\vec y), \label{mlett:a55} \\
& &\{\psi(x^{0}, \vec{x})\,,\,\pi_{\psi}(x^{0}, \vec{y})\}\,
=\,i\,\delta(\vec{x}-\vec{y}). \label{mlett:b55}
\end{eqnarray}
\end{mathletters}

The Coulomb gauge quantization of the MCS model in the presence of fermions
has so far been purely formal. To test its validity, we shall next use it to
compute the lowest order perturbative contribution to the electron-electron
elastic scattering amplitude (M\"oller scattering). Since the S-matrix is a
gauge invariant object, our result must coincide with that obtained for the
same process when working in covariant gauges\cite{Gi1}.

{}From the inspection of (\ref{56}) follows that the contributions of order
($e^{2}/\theta$) to the above mentioned amplitude, from now on referred to as
$R^{(2)}$, can be grouped into four different kind of terms

\begin{equation}
\label{58}
R^{(2)}\,=\,\sum_{\alpha=1}^{4}R^{(2)}_{\alpha},
\end{equation}

\noindent
where

\begin{mathletters}
\label{59}
\begin{eqnarray}
R^{(2)}_1 & = & -\frac {e^2}{2}(\gamma ^k)_{ab}(\gamma ^j)_{cd}
\int d^3x \int d^3y
\langle \Phi _f\vert T\left\{ :\bar \psi _a(x)\psi _b(x)A^k_T(x): \right.
\nonumber \\ & \times & \left.  :\bar \psi _c
(y)\psi _d (y)A^j_T(y):\right\}\vert \Phi _i\rangle\,, \label{mlett:a59} \\
R^{(2)}_2 & = & \frac {ie^2}{2}(\gamma ^0)_{ab}(\gamma ^0)_{cd}
\int d^3x\int d^3y
\delta (x^0-y^0)G_c(\vec x-\vec y) \nonumber \\
& \times & \langle \Phi _f\vert :\bar \psi _a (x)
\psi _b(x)\bar \psi _c(y) \psi _d(y):\vert \Phi _i\rangle\,,
\label{mlett:b59} \\
R^{(2)}_3 & = & -\frac {e^2}{2}\theta (\gamma ^k)_{ab}(\gamma ^0)_{cd}
\int d^3x \int d^3y
\langle \Phi _f\vert T\left\{ :\bar \psi _a(x)\psi _b(x)A^k_T(x): \right.
\nonumber  \\
& \times & \left.  :\epsilon
^{jm}A^m_T(y)\int d^3z \delta (y^0-z^0)\partial ^j_yG(\vec y-\vec z)
\bar \psi _c(z) \psi _d(z):\right\}\vert \Phi _i\rangle\,,
\label{mlett:c59} \\
R^{(2)}_4 & = & -\frac {e^2}{2}\theta ^2\epsilon ^{kl} \epsilon ^{jm}
(\gamma ^0)_{ab} (\gamma ^0)_{cd} \nonumber \\
& \times & \int d^3x \int d^3y
\langle \Phi _f\vert T\left\{ :A^l_T(x)\int d^3z_1 \delta (x^0-z^0)\partial
^k_x G_c(\vec x-\vec z_1)\bar \psi _a(z_1)\psi _b(z_1): \right.
\nonumber \\ & \times & \left. :A^m_T(y)
\int d^3z_2 \delta (y^0-z^0)\partial ^j_yG_c(\vec y-\vec z_2)\bar
\psi _c(z_2)\psi _d(z_2):\right\} \vert \Phi _i\rangle\,. \label{mlett:d59}
\end{eqnarray}
\end{mathletters}

\noindent
Here, $T$ is the chronological ordering operator and lower case latin
letters from the beginning of the alphabet are spin indices running
from $1$ to $2$. Also, $G_{c}(\vec{x}-\vec{y})$ is the Coulomb Green function
which, by definition, verifies $\nabla^{2}_{x} G_{c}(\vec{x}-\vec{y}) =
\delta(\vec{x}-\vec{y})$, while $|\Phi_{i} \rangle$ and $|\Phi_{f} \rangle$
denote the initial and final state of the reaction, respectively.

For the case under analysis, both $|\Phi_{i} \rangle$ and $|\Phi_{f} \rangle $
are two-electron states. Fermion states obeying the free Dirac equation in
2+1-dimensions were explicitly constructed in Ref.\cite{Gi1}, where the
notation $v^{(-)}(\vec{p})\, (\bar{v}^{(+)}(\vec{p}))$ was employed to
designate the two-component spinor describing a free electron of two-momentum
$\vec{p}$, energy $p^{0}=+({\vec{p}}^{\,2}+m^{2})^{1/2}$ and spin $s= m/|m|$ in
the initial (final) state. The plane wave expansion of the free fermionic
operators $\psi$ and $\bar{\psi}$ in terms of these spinors and of the
corresponding creation and anhilation operators goes as usual.

In terms of the initial ($p_{1}, p_{2}$) and final momenta
($p_{1}^{\prime}, p_{2}^{\prime}$), the partial amplitudes in (\ref{59}) are
found to read

\begin{mathletters}
\label{501}
\begin{eqnarray}
R^{(2)}_{1} & = & \frac {1}{2\pi}\delta^{(3)}(p_1'+p_2'-p_1-p_2)
\nonumber \\
& \times & \left\{ [\bar v^{(+)} (\vec{p}_1^{\,\prime})
(ie\gamma ^l)v^{(-)}(\vec p_1)][\bar v^{(+)}(\vec{p}_2^{\,\prime})
(ie\gamma^j)v^{(-)}(\vec p_2)] D^{lj}(k)\,- \, p_1^{\,\prime}
\longleftrightarrow p_2^{\,\prime} \right\}, \label{mlett:a501} \\
R^{(2)}_{2} & = & \frac{1}{2\pi}\delta ^{(3)}(p_1'+p_2'-p_1-p_2)
\nonumber \\
& \times & \left\{ [\bar v^{(+)}(\vec p_1^{\,\prime}) (ie\gamma ^0)
v^{(-)}(\vec p_1)]
[ \bar v^{(+)}(\vec p_2^{\,\prime})(ie\gamma ^0)v^{(-)}(\vec p_2)]
\frac {i}{\vert{\vec{k}}\vert ^2}\,-\, p_1^{\,\prime}
\longleftrightarrow p_2^{\,\prime} \right\}, \label{mlett:b501} \\
R^{(2)}_{3} & = & -\frac{\theta}{2\pi}\delta^{(3)}(p_1'+p_2'-p_1-p_2)
\nonumber \\
& \times & \left\{ [\bar v^{(+)}(\vec p_1^{\,\prime}) (ie\gamma ^l)
v^{(-)}(\vec p_1)]
[ \bar v^{(+)}(\vec p_2^{\,\prime})(ie\gamma ^0)v^{(-)}(\vec p_2)]
\Gamma ^l(k) \right.
\nonumber \\
& \times & \left. [\bar v^{(+)}(\vec p_1^{\,\prime}) (ie\gamma ^0)
v^{(-)}(\vec p_1)]
[ \bar v^{(+)}(\vec p_2^{\,\prime})(ie\gamma ^l)v^{(-)}(\vec p_2)]
\Gamma ^l(-k)\,
- \,p_1^{\,\prime} \longleftrightarrow p_2^ {\,\prime} \right\},
\label{mlett:c501} \\
R^{(2)}_{4} & = & \frac{\theta^{2}}{2\pi}\delta ^{(3)}(p_1'+p_2'-p_1-p_2)
\nonumber \\
& \times & \left\{ [\bar v^{(+)} (\vec p_1^{\,\prime})(ie\gamma ^0)
v^{(-)}(\vec p_1)]
[ \bar v^{(+)}(\vec p_2^{\,\prime})(ie\gamma ^0)v^{(-)}(\vec p_2)]
\Lambda(k)\,
- \,\,p_1^{\,\prime} \longleftrightarrow p_2^ {\,\prime} \right\},
\label{mlett:d501}
\end{eqnarray}
\end{mathletters}

\noindent
where\footnote{Our convention for the Fourier integral representation is
$f(x)= 1/(2\pi)^{3}\int d^{3}k f(k) \exp(ik\cdot x)$}

\begin{mathletters}
\label{502}
\begin{eqnarray}
&& D^{lj}(k)\,=\,\frac {i}{k^2-\theta ^2+i\epsilon}\left(\delta
^{jl}-\frac{k^jk^l}
{\vert \vec k\vert ^2}\right), \label{mlett:a502} \\
&& \Gamma ^l(k)\,=\,\frac {\epsilon ^{lj}k^j}{(k^2-\theta ^2+i\epsilon)}
\frac {1}{\vert \vec k\vert ^2}, \label{mlett:b502} \\
&& \Lambda (k)\,=\,\frac{i}{\vert \vec k\vert ^2(k^2-\theta ^2+i\epsilon )}
\label{mlett:c502},
\end{eqnarray}
\end{mathletters}

\noindent
and

\begin{equation}
\label{503}
k \equiv  p_1'-p_1\,=\,-(p_2'-p_2),
\end{equation}

\noindent
is the momentum transfer. By going back with (\ref{501}) into (\ref{58}) and
after taking into account (\ref{502}) one arrives at

\begin{eqnarray}
\label{504}
R^{(2)}\,& = &\,\left( -\frac{e^2}{2\pi}\right) \delta^{(3)}(p_1'+p_2'-p_1-p_2)
D_{\mu \nu}(k) \nonumber \\
& \times & \left\{ [\bar v^{(+)}(\vec p_1^{\,\prime})
\gamma^{\mu} v^{(-)}(\vec p_1)]\,[\bar v^{(+)}(\vec p_2^{\,\prime})
\gamma^{\nu} v^{(-)}(\vec p_2)]
- p_1^{\,\prime} \longleftrightarrow p_2^ {\,\prime} \right\}\,,
\end{eqnarray}

\noindent
where

\begin{equation}
\label{505}
D_{\mu \nu}(k) = -\frac{i}{k^2-\theta ^2+i\epsilon }\left[ g_{\mu \nu} +
i\theta \epsilon _{\mu \nu \rho}\frac{\bar k^{\rho}}{\vert \vec k\vert ^2}
\right]
\end{equation}

\noindent
and ${\bar{k}}^{\rho} = (0, \vec{k})$. In the literature\cite{Ja,Lo},
the effective Coulomb gauge propagator in Eq.(\ref{505}) has been referred to
as the infrared safe propagator. We stress that, unlike the case in
electrodynamics ($\theta=0$), $D_{\mu \nu}(k)$ does not turn to be a covariant
object. One also learns from Eq.(\ref{504}) that, when $D_{\mu \nu}(k)$ is used
as the Coulomb gauge photon propagator, one is to replace all the
non-covariant terms in $H_I$ by the standard field-current minimal interaction.

As for the equivalence between the Coulomb and the covariant gauges, we start
by recalling that $R^{(2)}$ is a Lorentz invariant object and, then,
(\ref{504}) can be evaluated in any Lorentz frame. In the center of mass frame,
the energy transfer,

\[
k^{0} \equiv  p_{1}^{\prime\,0} - p_1^{0}\,=\,0,
\]

\noindent
vanishes and, whence, nothing changes if one replaces in the right hand side of
the last mentioned equation

\begin{eqnarray}
\bar{k}^{\rho} & \longrightarrow & k^{\rho} \nonumber \\
|\vec{k}^{2}| & \longrightarrow & -\, k^{2} \nonumber\,.
\end{eqnarray}

\noindent
After these replacements, Eq.(\ref{504}) can be written

\begin{eqnarray}
\label{506}
R^{(2)}\,& = &\,\left( -\frac{e^2}{2\pi}\right) \delta^{(3)}(p_1'+p_2'-p_1-p_2)
D_{\mu \nu}^{L}(k) \nonumber \\
& \times & \left\{ [\bar v^{(+)}(\vec p_1^{\,\prime})
\gamma^{\mu} v^{(-)}(\vec p_1)]\,[\bar v^{(+)}(\vec p_2^{\,\prime})
\gamma^{\nu} v^{(-)}(\vec p_2)]
- p_1^{\,\prime} \longleftrightarrow p_2^ {\,\prime} \right\}
\end{eqnarray}

\noindent
where

\begin{equation}
\label{507}
D_{\mu \nu}^{L}(k) = -\frac{i}{k^2-\theta ^2+ie}\left[ g_{\mu \nu} -
i\theta \epsilon_{\mu \nu \rho}\frac{k^{\rho}}{k^2}
\right]
\end{equation}

\noindent
is the covariant Landau gauge propagator. Again, Lorentz invariance secures
that (\ref{506}) holds in all Lorentz frames. The equivalence between the
Coulomb and the Landau gauge is by now established. To extend this
equivalence to other covariant gauges, it is enough to observe that
$D_{\mu \nu}^{L}(k)$ in Eq.(\ref{506})is contracted into conserved currents
and, hence, terms proportional $k^{\mu}$ can be added at will.

\section{M\"oller Scattering to All Orders of Perturbation Theory}
\label{sec:level6}

As it was shown in the previous Section, the Coulomb gauge Feynman rules are
those of the covariant gauges with the covariant propagator replaced by the
infrared safe propagator (\ref{505}). This Section is dedicated to
demonstrate that this argument alone suffices to secure that, in any order of
perturbation theory, the M\"oller scattering amplitudes computed in the Coulomb
and in the covariant Landau gauge are indeed the same. The basic observation is
that the difference between the Landau and the infrared safe propagator,
given at Eqs.(\ref{507}) and (\ref{505}), respectively, is of the form

\begin{equation}
\label{61}
D_{\mu \nu}^{L}(k)\,-\,D_{\mu \nu}(k)\,=\,k_{\mu} G_{\nu}(k)\,-\,k_{\nu}
G_{\mu}(k)\,,
\end{equation}

\noindent
where

\begin{equation}
\label{62}
G_{\nu}(k)\,\equiv \, \theta \frac{\epsilon_{\nu 0 j}}
{k^{2} (k^{2} - \theta^{2} + i\epsilon)}\, \frac{k^{j} k^{0}}{{\vec{k}}^{2}}\,.
\end{equation}

\noindent
Consequently, the purported proof of equivalence has now been reduced to show
that if in all graphs, of a given order in perturbation theory, one of the
photon propagators is replaced by the right hand side of (\ref{61}), the sum of
these modified graphs vanish.

To see how this come about we examine the generic diagram in Fig.~\ref{fig01},
where an internal photon line has been singled out. After performing the above
mentioned replacement the graph decomposes into the sum of two pieces, each
containing either $k_{\mu} G_{\nu}(k)$ or $k_{\nu} G_{\mu}(k)$. We
study first the part containing $k_{\mu} G_{\nu}(k)$, hereafter referred to as
$P_{\mu \nu}$. We furthermore assume that the incoming and outgoing fermion
lines meeting at the vertex $\mu$ carry momentum $p$ and $p+k$, respectively.
By using

\begin{equation}
\label{63}
\gamma \cdot k \,=\,[\gamma \cdot (p + k) - m]\,
- \,(\gamma \cdot p - m)\,,
\end{equation}

\noindent
one gets

\begin{equation}
\label{64}
\frac{1}{\gamma \cdot p - m}\,\, \gamma \cdot k \,\,\frac{1}
{\gamma \cdot (p + k) - m}\,=\,\frac{1}{\gamma \cdot p - m}
\,-\,\frac{1}{\gamma \cdot (p + k) - m}.
\end{equation}

\noindent
Thus, the fermion lines meeting at the vertex under analysis become amputated,
one at the time. This in turn implies that $P_{\mu \nu}$ splits into the sum
of two contracted subgraphs.

If the amputated line was an external fermion line, the corresponding
contracted subgraph vanishes due to the on shell condition.

If the amputated line was an internal fermion line we must still distinguish
two possibilities. First, the amputated line did not link directly the vertices
$\mu$ and $\nu$. When this is the case, the contribution to the amplitude
arising from this contracted subgraph is cancelled by the contribution made
by a topollogically equivalent contracted subgraph, originating from another
diagram. Secondly, the amputated line did link the vertices $\mu$ and $\nu$.
Then, it is clear that the contracted subgraph contains the tadpole integral

\[
\int d^3 k G_{\nu}(k)
\]

\noindent
which can be seen to vanish due to symmetry requirements. Through similar
arguments one shows that $P_{\nu \mu}$ also vanishes.

The proof of gauge independence for the M\"oller scattering amplitude in the
MCS theory is now complete. It greatly helps to establish the reliability of
the quantization of the MCS model in the Coulomb gauge.

\section{Conclusions}
\label{sec:level7}

We started in this work by looking for the Coulomb gauge polarization vector of
the free MCS theory. We found impossible to determine such
vector by working only within the Coulomb gauge. The reason for this being
clear, the MCS theory is a gauge theory whose gauge particle is massive
when formulated in the Coulomb gauge. Hence, one can think of a frame of
reference where this quanta is at rest. In such frame ($\vec{k}=0$) the
Coulomb condition $\vec{k}\cdot \vec{\epsilon}(\vec{k})=0 $ is not longer
operative. The
way out from the trouble consisted in finding first the polarization vectors
of the Landau gauge, where there exist massive and massless gauge particles.
The massless excitation is a gauge artefact that dissapears when gauge
transformed into the Coulomb gauge. The polarization vector associated with
the massive excitation provides, after being gauge transformed,
the Coulomb gauge polarization vector, which turned out to be free of
ambiguities. In 2+1-dimensions the Coulomb gauge appears to be as
respectable as it is in 3+1-dimensions, in the sense that it only allows for
physical excitations.

We constructed, afterwards, a set of quantum densities obeying the
Dirac-Schwinger
algebra. These densities were taken as the Poincar\'e densities and served to
build the Poincar\'e generators. Particular attention was dedicated to the
generator of spatial rotations. It was shown that the spin of the massive
gauge particles originates from the highly unconventional mathematical
structure of the Coulomb gauge polarization vector.

The MCS gauge particles were then allow to interact with charged fermions.
A practical test of the Coulomb gauge Feynman rules for the interacting theory
was carried out in connection with the M\"oller scattering amplitude. As
demanded by gauge invariance, the result was shown to agree, to all orders in
perturbation theory, with that obtained by using covariant Feynman rules.
This proof was based on the observation that the Coulomb gauge Feynman rules
can be phrased in terms of an effective photon propagator, the infrared safe
propagator\cite{Ja,Lo}, which allows for the replacement of all terms in the
interaction Hamiltonian (see Eq.(\ref{56})) by the standard field-current
minimal
interaction.

To summarize, we believe to have contributed to establish that the quantization
of the MCS in the Coulomb gauge is free of inconsistencies and can be
systematically carried out.

\begin{figure}
\caption{A generic Feynman graph contributing to the M\"oller scattering
amplitude. An internal photon line has been singled out. The box stands for the
rest of the diagram.}
\label{fig01}
\end{figure}


\begin{references}

\bibitem[*]{byline} Supported in part by Conselho Nacional de Desenvolvimento
Cient\'{\i}fico e Tecnol\'ogico, CNPq, Brazil.

\bibitem {Ja} S. Deser, R. Jackiw and S. Templeton, Ann. Phys. (NY) {\bf 140}
(1982), 372 .

\bibitem {Sch} J. F. Schonfeld, Nucl. Phys. {\bf B185} (1981), 157 .

\bibitem {Ha} C. R. Hagen, Ann. Phys. (NY) {\bf 157} (1984), 342 .

\bibitem {Fo} A. Foerster and H. O. Girotti, Statistical transmutation in
(2+1)-dimensions, in J. J. Giambiagi Festschrift, eds. H. Falomir, R. E. Gamboa
Sarav\'{\i}, P. Leal Ferreira and F. A. Schaposnik (World Scientific,
Singapore, 1990) p.161; Phys. Lett {\bf B230} (1989), 83;
Nucl. Phys. {\bf B342} (1990), 680 .

\bibitem {Di}  P. A. M. Dirac, {\it Lectures on Quantum Mechanics},
Belfer Graduate School of Science, Yeshiva University (New York,
1964).

\bibitem {Fr} E. S. Fradkin and G. A. Vilkovisky, CERN preprint TH
2332 (1977), unpublished.

\bibitem {Su} K. Sundermeyer, {\it Constrained Dynamics} (Springer-Verlag,
Berlin, 1982).

\bibitem {Gi} H. O. Girotti, {\it Classical and quantum dynamics of
constrained systems}, lectures in Proc. of the $ V^{th} $ Jorge
Andre Swieca Summer School, O. J. P. Eboli, M. Gomes and A.
Santoro editors (World Scientific, Singapore, 1990) p1-77.

\bibitem {Sch1} J. Schwinger, Phys. Rev. {\bf 127} (1962), 324. Also see
C. Itzykson and J-B. Zuber, {\it Quantum Field Theory} (McGraw-Hill,
Singapore, 1985).

\bibitem {Gi1} H. O. Girotti, M. Gomes and A. J. da Silva,
Phys Lett. {\bf B274} (1992), 357.

\bibitem {Gi2} H. O. Girotti, M. Gomes, J. L. deLyra, R. S. Mendes,
J. R. S. Nascimento and A. J. da Silva, Phys. Rev. Lett. {\bf 69} (1992),
2623.

\bibitem {Lo} O. Bergman and G. Lozano, Ann. Phys (NY) {\bf 229} (1994), 229.

\bibitem {Mo} Ya. I. Kogan and A. Yu. Morozov, Sov. Phys. JETP {\bf 61}
(1985), 1.

\bibitem {Jap1} T. Kimura, Prog. Theor. Phys. {\bf 81} (1989), 1109.

\bibitem {Jap2} N. Imai, K. Ishikawa and I. Tanaka, Prog. Theor. Phys {\bf 81}
(1989), 758. Also see N. Nakanishi, Int. J. Mod. Phys. {\bf A4} (1989), 1055.

\bibitem {Bo} D. Boyanovsky, R. Blankenbecler and R. Yahalom, Nucl. Phys
{\bf B270} (1986), 483.

\bibitem {Bi} B. Binegar, J. Math. Phys. {\bf 23} (1982), 1511.
\end{references}
\end{document}